\documentclass[a4paper,12pt]{article}
\usepackage{hyperref}
\usepackage[english]{babel}
\usepackage[T1]{fontenc}
\usepackage[utf8]{inputenc}
\usepackage{amsfonts}
\usepackage{amssymb}
\usepackage{amsmath}
\usepackage{graphicx}
\usepackage{cite}
\usepackage{epsfig}
\usepackage{psfrag}
\usepackage{tikz}
\usetikzlibrary{matrix,arrows,decorations.pathmorphing,positioning} 
\usepackage{url}
\usepackage{amsthm}
\usepackage{mathpartir}
\usepackage{enumerate}
\usepackage{tikz-cd}

\newtheorem{definition}{Definition}[section] 
\newtheorem{lemma}{Lemma}[section]  
 
\newtheorem{remark}{Remark}[section]

\newcommand{\angled}[1]{\left< #1 \right>}

\date{}

\title{\bf On the Operator Origins of Classical and Quantum Wave Functions}

\author{Xerxes D. Arsiwalla$^{1, 2, }$\footnote{Corresponding Author:  \url{x.d.arsiwalla@gmail.com}}  { }  David Chester$^{3, }$\footnote{\url{DavidC@quantumgravityresearch.org}}   { }  Louis H. Kauffman$^{4, }$\footnote{\url{loukau@gmail.com}}  \\  
{}  \\
{\it \small $^{1}$Pompeu Fabra University, Barcelona, Spain}\\ 
{\it \small $^{2}$Wolfram Research, USA}\\
{\it \small $^{3}$Quantum Gravity Research, USA}\\
{\it \small $^{4}$University of Illinois at Chicago, USA}    
}

\begin{document}
\maketitle

\begin{abstract}

We investigate operator algebraic origins of the classical Koopman-von Neumann wave function $\psi_{KvN}$ as well as the quantum mechanical one  $\psi_{QM}$. We introduce a formalism of Operator Mechanics (OM) based on a noncommutative Poisson, symplectic and noncommutative differential structures. OM serves as a pre-quantum algebra from which algebraic structures relevant to real-world  classical and quantum mechanics follow. In particular, $\psi_{KvN}$ and $\psi_{QM}$  are both consequences of this pre-quantum formalism. No a priori Hilbert space is needed. OM admits an algebraic notion of operator expectation values without invoking  states. A phase space bundle ${\cal E}$ follows from this. $\psi_{KvN}$ and $\psi_{QM}$ are shown to be sections in ${\cal E}$. The  difference between  $\psi_{KvN}$ and $\psi_{QM}$ originates from a quantization map interpreted as  "twisting" of sections over  ${\cal E}$.  We also show that the Schr\"{o}dinger equation is obtained from the Koopman-von Neumann equation. What this suggests is that neither the Schr\"{o}dinger equation nor the quantum wave function are fundamental structures. Rather, they both originate from a pre-quantum operator algebra. Finally, we comment on how entanglement between these operators suggests emergence of space; and possible extensions of this formalism to field theories. 

\end{abstract}

\vspace{2pc}
{\it Keywords}: Operator Algebras, Noncommutative Differential Calculus,  Koopman-von Neumann Mechanics, Pre-Hilbert Spaces, Operator Entanglement.

\clearpage

\tableofcontents

%\clearpage

\section{Introduction}

Many traditional formulations of quantum mechanics typically assume a Hilbert space of states ${\cal H}$ upon which one admits an algebra of linear operators over ${\cal H}$. Observables quantities correspond to Hermitian operators in this algebra and measurements correspond to expectation values of these operators with respect to a suitably defined density operator. In other words, states take precedence over operators with the Hilbert space of states as the primary structure upon which everything else is defined or constructed.  
%\footnote{These statements hold true irrespective of whether one is working in the Schr\"{o}dinger, Heisenberg or Dirac picture.}.  
There certainly do exist approaches to quantum mechanics where operators take precedence over states. An example being `Categorical Quantum Mechanics' (CQM) \cite{abramsky2009categorical}, where one typically considers the compact closed monoidal category of finite-dimensional Hilbert spaces (referred to as {\bf FHilb}). The objects of  {\bf FHilb}  are finite-dimensional Hilbert spaces used to describe `states', and whose morphisms are linear operators used to describe `processes'.   

Another related example, where operators take precedence, is Connes' Noncommutative Geometry (NCG), as an attempt to describe the quantum geometry of spaces using noncommutative coordinates   \cite{connes1985non}.  In this case, the underlying base structure of the formalism is  a  Fredholm module or a spectral triple $({\cal A}, {\cal H}, F)$, which consists  of a representation of a noncommutative ${\cal C}^\star$ operator algebra on a Hilbert space together with an unbounded self-adjoint operator. However, as mentioned, even in these examples, a Hilbert space of states is a crucial part of the initial structure of the formalism. In other words, this dichotomy between states and operators is very much one between geometry and algebra respectively. The original motivation for the NCG program was to investigate whether one can indeed construct  geometric data of a base manifold ${\cal M}$ from a noncommutative operator algebra. The cotangent bundle ${\cal T}^\star{\cal M}$ of  ${\cal M}$ or rather its configuration space provides the support of the Hilbert space of $L^2$-integrable functions. Observables then correspond to Hermitian operators on this Hilbert space. 

On the other hand, what we would like to investigate in this work is whether one can solely work with an abstract operator algebra, independent of any specific representation or a priori Hilbert space, and rather construct the space of states from these operators alone. In other words, absolutely without taking recourse to any geometric overheads.  A purely abstract algebraic formalism where the space of states itself is obtained as a derived entity, would arguably be a promising candidate for a pre-quantum theory, or rather for pre-quantum geometry.  What can be useful with such pre-quantum structures is that they allow for background-independent formulations of potential quantum gravity models. In particular, these methods may help concretize pregeometric approaches to quantum theory, of the kind advocated by Wheeler  \cite{wheeler1980pregeometry}.       

In contrast to some of the above-mentioned programs, the objectives of our work here are modest. As a first step towards a pre-quantum theory, we introduce a formalism of Operator Mechanics (OM), based on a Lie algebra of canonically conjugate operators. This algebra is equipped with a noncommutative Poisson structure, a symplectic structure and a noncommutative differential structure.  The OM framework introduced here builds upon previous work on noncommutative geometry discussed in   \cite{kauffman1998noncommutativity,kauffman2004non,kauffman2018non,kauffman2022calculus}. OM is presented as a formal operator algebra and this framework does not rely on any specific representation theory of the algebra, nor its action on any Hilbert space. In this sense, OM assumes far less than the  noncommutative geometry formalism of Connes', whose starting point is a spectral triple.  In OM, we recover the operator version of Hamilton's classical equations of motion. This follows simply as a consequence of the symplectic structure. An important ingredient of OM that we introduce here is a definition of a density operator defined using the symplectic structure of the algebra (without alluding to any notion of states). This allows us to compute expectation values of operators without invoking any notion of wave functions or states. In the OM  framework, wave functions are not  fundamental entities, but are derived from structures built upon the operator algebra\footnote{There also exist  formulations of quantum mechanics that completely do away with wave functions. This is the `phase space formulation' of Groenewold and Moyal \cite{groenewold1946principles,moyal1949quantum}, which is based on quasi-probability distributions over phase space. In contrast, in OM we only assume an abstract operator algebra as our starting point and motivate the construction of phase space and the algebra of functions over it.}.

We will investigate both, the classical Koopman-von Neumann's $\psi_{KvN}$  \cite{koopman1931hamiltonian,neumann1932operatorenmethode} and the quantum mechanical wave function $\psi_{QM}$.  We do not assume any a priori Hilbert space of states. OM admits an algebraic notion of operator expectation values without invoking wave functions or states. A phase space bundle ${\cal E}$ follows from this. Both $\psi_{KvN}$ and $\psi_{QM}$ are shown to be sections in ${\cal E}$. The  difference between  $\psi_{KvN}$ and $\psi_{QM}$ appears in the form of a quantization scheme interpreted as  "twisting" of sections over  ${\cal E}$.   Implementing  this quantization map, we will also show that the Schr\"{o}dinger equation itself  can be obtained from the Koopman-von Neumann equation. What this suggests is that neither the Schr\"{o}dinger equation nor the quantum mechanical wave function are fundamental structures. Rather, these are built upon a pre-quantum operator algebra and ensuing algebraic homomorphisms.   In this sense, the OM formalism presented here serves as a candidate for a  pre-quantum operator algebra. 

The outline of this paper is as follows: In Section 2, we introduce our definitions  of the OM formalism. In Section 3, we discuss the operator equations of motion resulting in OM and its pre-quantum nature. In Section 4, we introduce a new density operator and show that it satisfies the usual requirements of a density operator; we also introduce a new operator expectation value map. In Section 5, we discuss structures leading to wave functions, observables and expectation values of the second kind. In Section 6,  we discuss the pre-quantum origins of Ehrenfest's theorem. In Section 7,  we show how the Schr\"{o}dinger equation itself  follows from the Koopman-von Neumann equation via quantization.  In Section 8, we discuss further implications of pre-quantum mechanics and conclude, mentioning connections to matrix models, noncommutative geometry, emergence of space from entanglement and other category-theoretic approaches.

\section{Definitions of Operator Mechanics (OM) }

We now introduce our framework of Operator Mechanics (OM).  The mathematical formulation of OM is based upon a noncommutative Poisson algebra \cite{crawley2007noncommutative,van2008double,bocklandt2002necklace,xu1994noncommutative}, a symplectic structure and a noncommutative differential structure. 

Recall that a commutative Poisson algebra is a commutative (unital) algebra  (${\cal L}, \, +, \,  \cdot$) with  an associative product `$\cdot$', and at the same time also a Lie algebra  (${\cal L},   \, +, \,   [ \, -  ,  - \, ]$)  with a non-associative product  $ [ \, -  ,  - \, ] : {\cal L} \times {\cal L} \to {\cal L}$, that is bilinear and anti-commutative. Furthermore, this algebra satisfies the Jacobi identity  and the  Leibniz rule respectively:  
\begin{eqnarray}
\,  [f, \, [ g,  h] ]    &=&   [ [f, g], \, h]  +  [g,  \, [f,  h] ]   \\
\,  [f, \, g \cdot  h] &=&  \,\,   [ f,  g] \cdot  h  \, + \,  g \cdot [f,  h]  
\end{eqnarray}
where $f$, $g$, $h$ belong to ${\cal L}$.  The Jacobi identity itself can be thought of as a Leibniz rule with respect to the Lie product.  Notice that if one defines a derivative $D (g) = [f, g]$ using Lie brackets and a given $f$ in this algebra, then the Leibniz rule stated above, yields the usual product rule for derivatives. On the other hand, without the product `$\cdot$', one simply recovers the definition of a Lie algebra.   The archetypical example of a commutative Poisson algebra is the algebra of  ${\cal C}^{\infty}$ functions with the Poisson bracket as the Lie bracket. This forms the backbone of classical mechanics. 

To define a noncommutative Poisson algebra, the product `$\cdot$' is now replaced by a noncommutative  associative product.  ${\cal L}$ can be an operator ring or an ${\cal R}$-module and the bracket 
\begin{eqnarray}
 [A,  B] = A \cdot B - B \cdot  A
 \label{lb}
\end{eqnarray}
realizes a noncommutative Poisson structure\footnote{For notational simplification we will hereon drop the $\cdot$  and only express  products $A \cdot B$ as $A \, B$.}.  Only few other instances of noncommutative Poisson structures are known. These have been found on moduli spaces classifying semi-simple modules  \cite{crawley2007noncommutative,van2008double}.    Examples of such systems include path algebras of doubled quivers, pre-projective algebras and multiplicative pre-projective algebras discussed in  \cite{crawley2007noncommutative,van2008double}.  In what follows, we will remain largely agnostic to the choice of Lie brackets. In OM, $[ \, -  ,  - \, ]$ will denote a generic Lie bracket, whose form could be as in Eq.~(\ref{lb}) or one  appearing in aforementioned cases of path algebras of doubled quivers and pre-projective algebras that have noncommutative Poisson structures given by the necklace Lie algebra   \cite{bocklandt2002necklace,crawley2007noncommutative,van2008double}. 

Furthermore, in addition to this noncommutative Poisson structure we will include a symplectic structure on (${\cal L}, \, +, \,  \cdot$).  Let this algebra be generated by operators $\{ X^i \}$ and $\{ P_i \}$, for $1 \leq i \leq n$, plus the multiplicative identity. These generators satisfy the following canonical commutation relations:
%  Within ${\cal A}$ one can study various sub-algebras endowed with specific structures. A particularly useful sub-algebra  turns out to be a Lie algebra ${\cal L} \subseteq {\cal A}$ defined by the bracket  
%\begin{eqnarray}
% [ \, -  ,  - \, ] : {\cal L} \times {\cal L} \to {\cal L} 
%\end{eqnarray}
%This bracket can be a generic Lie bracket and the requirements for setting up OM are independent of the specific representation of this bracket. The only requirements being that it satisfies (i) the axioms of a Lie algebra, that is, it be  bilinear, anti-symmetric and satisfies the Jacobi identity; and (ii) the Leibniz rule. 
\begin{eqnarray}
[X^i,  P_j]  &=&    \kappa  \;  \delta^i_j  \label{3}  \\   
\, [X^i,  X^j]  &=&  0   \label{4}  \\ 
\, [P_i,  P_j]  &=&  0  \label{5}
\end{eqnarray} 
where $\kappa$ is an arbitrary constant at this point (which need not be the $i \, \hbar$ of quantum mechanics, but may be associated to a more fundamental theory). 

Let us make a small remark on the dimensions of $\kappa$, which in general will depend on specific representations of the Lie bracket that one works with. When the Lie bracket  $[ \, -  ,  - \, ]$ is chosen as in   Eq.~(\ref{lb})  above, then $\kappa$ has dimensions M\,L$^2$T$^{-1}$. On the other hand if one replaces  (${\cal L}, \, +, \,  \cdot$)  by a commutative algebra, then the Lie bracket is the Poisson bracket. In that case, $\kappa$  would equal $1$ and be dimensionless. 

%Extending beyond the fully commuting case, a slightly more interesting Lie algebra with noncommutative operators can be constructed using the following bracket:
%\begin{eqnarray}
% [A,  B] = \sum_{i = 1}^{n} \left( \frac{\partial A}{\partial X^i}  \frac{\partial B}{\partial P_i}  -  \frac{\partial A}{\partial P_i}  \frac{\partial B}{\partial X^i}  \right)  - \sum_{i = 1}^{n}    \left( \frac{\partial B}{\partial X^i}  \frac{\partial A}{\partial P_i}  -  \frac{\partial B}{\partial P_i}  \frac{\partial A}{\partial X^i}  \right)   
%\label{npb}
%\end{eqnarray}
%where $A$ and $B$ are non-commutative operators functions depending on $X^i$ and $P_i$ and the derivatives are those defined in the context of  Weyl calculus in noncommutative analysis (see \cite{jefferies2004spectral}).  
%For this choice of bracket, $\kappa$  equals $2$  and is once again  dimensionless. This can be seen by evaluating Eq.~(\ref{npb}) for $X^i$ and $P_i$ gives $[X^i,  P_j] = 2$, which gives $\kappa = 2$. Hence, we exactly recover  eqs. (\ref{3}), (\ref{4}) and (\ref{5}).  Eq.~(\ref{npb}) is yet another example of a bracket that satisfies the canonical commutation relations.  This is a  noncommutative generalization of the classical Poisson bracket, and in an  upcoming work we shall explore its properties and potential connections to the current work.  

The symplectic structure underlying the canonical commutation relations in eqs. (\ref{3}), (\ref{4}) and (\ref{5}) can be seen by packaging operators $X^i$ and $P_i$ into "vectors" (or rather elements of an operator module)  
\begin{eqnarray}  
 V_i = ( X^i, \, P_i )^T
\end{eqnarray}
The components of this object $V_i^a$ for $a = \{ 1, 2 \}$ then satisfy the following:
\begin{eqnarray}  
\frac{1}{\kappa}  \, [V_i^a,  V_j^b]  &=&    \omega^{a b}  \;  \delta_{i j}
\end{eqnarray}
where $\omega^{a b}$ is  the symplectic form with matrix representation
\begin{eqnarray}  
    \omega  &=& \left( \begin{matrix}  0 & 1  \\  -1 &  0  \end{matrix} \right) 
\end{eqnarray}

%Corresponding to the bracket in Eq.~(\ref{lb}), we see that Eq.~(\ref{3})  can also be thought of as an inner product between vectors in phase space, induced by the symplectic form
%\begin{eqnarray}
%  \begin{pmatrix}  X^i  &  P_i   \end{pmatrix}   \begin{pmatrix}  0 &  \delta^i_j  \\  -     \delta^i_j   &  0     \end{pmatrix}    \begin{pmatrix}  X^j  \\  P_j   \end{pmatrix}  \, =  \,  \kappa  \;  \delta^i_j     
%\end{eqnarray} 

Given a noncommutative Poisson and symplectic structure, OM admits the   following differential structure:   
\begin{definition}
Partial derivatives of operators $F \in {\cal L}$  are maps Der
\begin{eqnarray}
Der : {\cal L} \to  {\cal L} 
\end{eqnarray}
%where $E$ is an ${\cal L}$ bi-module, and 
where Der satisfies the Leibniz Rule: 
\begin{eqnarray}
Der (a \cdot b) = Der (a) \cdot b + a \cdot Der (b) 
\end{eqnarray}
With respect to $X^i$ and $P_i$, the generators of the algebra, these partial  derivatives are defined as: 
\begin{eqnarray}
\kappa  \, \frac{\partial F}{\partial X^i}  &=& [F,  P_i]   \label{dx}  \\
\kappa  \,  \frac{\partial F}{\partial P_i}  &=& [X^i,  F]  \label{dp}
\end{eqnarray}
\end{definition}

\begin{definition}
The time derivative of an operators $F \in {\cal L}$ is defined as a Lie product with a special operator $H$, which we call the Hamiltonian
\begin{eqnarray}
\kappa  \,  \frac{d F}{d t}  &=&  [F,  H]  
\label{dt}
\end{eqnarray}
\end{definition}
Note that this form of the time derivative is structurally analogous to the  Hamiltonian flow generated by a Lie algebra.  The brackets $[ \, -  ,  - \, ]$ in both the above definitions are generic Lie brackets, which algebraically capture  the noncommutative differential structure of OM.

\section{OM as a Pre-Quantum Algebra }

As detailed above, our formalism of OM  is the noncommutative analogue of the Poisson algebra (plus a symplectic structure).  The commutative case is defined over  ${\cal C}^{\infty}$ functions, whereas here we will work with a noncommutative operator algebra.  In OM, we take the conjugate variables $X^i$ and $P_i$ to be abstract operators (that is, those defined independent of an operator  representation or upon a Hilbert space) satisfying the canonical commutation relations. We  replace Poisson brackets with Lie brackets admitting the Leibniz rule over  noncommutative algebras.  %This generalizes a commutative algebra of functions with a noncommutative algebra of operators and ${\cal R}$-modules over operator rings. 
We will take the perspective that abstract operator algebras be deemed neither classical nor quantum entities as they carry insufficient structure to  completely specify either, yet they demonstrate certain properties referring to  both (as we shall now discuss).  Special realizations of ${\cal L}$ (examples may also include a sub-algebra involving a commutative ring or other noncommutative deformations of the Poisson bracket) with a given specification of the Lie bracket, and appropriate $\kappa$ values, may accordingly yield classical or quantum mechanical characteristics based on subsequent structures admissible over the operator algebra. In this sense, we refer to OM as a pre-quantum framework upon which the laws of classical and quantum mechanics arise following the identification of additional algebraic structures. This confronts us with an important fundamental question: Where does the wave function and   algebra of observables (in both, classical and quantum mechanics) come from?  In this work we investigate algebraic structures that may help shed light on this question. 

%Later we will see that particular deformations of $\kappa$ and more generally Clifford algebraic deformations of ${\cal L}$ will yields different classes of quantum theories in noncommutative geometry. 

We now show how the operator analogues of  Hamilton's equations of motion can be derived in our formalism.  A special case of this result was shown in  \cite{kauffman1998noncommutativity,kauffman2004non,kauffman2018non,kauffman2022calculus}  when one considers the commutator form of the Lie bracket as in Eq.~(\ref{lb}). Here we extend that result for any generic Lie bracket as follows: 
\begin{eqnarray}
\kappa  \,  \frac{d P_i}{d t}  &=&  [P_i,  H]  \, = \,  - [H, P_i]  \, = \,  -  \kappa  \,  \frac{\partial H}{\partial X^i}  \label{hameq1}  \\ 
\kappa  \,  \frac{d X^i}{d t}  &=&  [X^i,  H]  \, = \, \kappa  \,  \frac{\partial H}{\partial P_i}  
\label{hameq2}
\end{eqnarray}
where only the definitions of derivatives and axioms of a Lie algebra have been used, without alluding to any representation of the Lie bracket.  The   Hamilton's equations thus derived are also independent of the value of $\kappa$ (with the constant dropping off on both sides). 

From the above, it is also straightforward  to see how Newton's laws (in operator form) can be obtained in our  noncommutative OM formalism for any Hamiltonian quadratic in momentum.  Consider the Hamiltonian:  
\begin{eqnarray}
H = \frac{1}{2 m}  \sum_{i,j = 1}^n \eta^{ij} P_i P_j + V (\{ X^i \})
\label{1part0}
\end{eqnarray}
with mass $m$, $\eta^{ij}$ a flat Euclidean metric, and $V (\{ X^i \})$ an arbitrary operator-valued potential. Substituting $H$ in Hamilton's equations above gives  
\begin{eqnarray}
\frac{d P_i}{d t}  &=&  -  \frac{\partial V (X)}{\partial X^i}  \label{1part2}  \\ 
\frac{d X^i}{d t}  &=&  \frac{\eta^{ij}}{m}  P_j   \label{1part1}
\end{eqnarray} 
where the derivatives are the noncommutative differentials defined in Section 2 above. Putting the above equations together, we get the following operator-theoretic version of Newton's law of motion: 
\begin{eqnarray}
m \ddot{X^i} = { F^i}
\label{newtlaw}
\end{eqnarray}
where the force ${ F^i}$ is defined as the negative gradient of the potential $- \partial V (X) / \partial X^i$ (the partial derivative being an OM commutator)\footnote{The notation $X$ is simply shorthand for the $n$-tuple of components $\{ X^i \}$.}.

Note that Eqs. (\ref{hameq1}) - (\ref{newtlaw})  are all operator equations.  Naive application of  standard calculus or functional analysis in order to solve these equations of motion will not work. Instead, one has to work with spectral properties and admissible representations of these operators to find solutions.  This is because with noncommutativity, one is confronted with what is sometimes referred to as  "pointless geometry", where localized notions of coordinates on phase space are replaced by operators.   

On the one hand, the above noncommutative formalism seems to bear partial resemblance to classical mechanics; on the other hand it also resembles quantum mechanics.  The latter can be seen from the OM time evolution equation Eq.~(\ref{dt}), which has the form of Heisenberg's equation (for operators without explicit time dependence) when $\kappa$ is taken to be $i \, \hbar$.  However, by itself the algebra of OM is strictly-speaking neither classical nor quantum. It is an abstract algebraic structure defined independently of both, background spaces and state spaces (classical or quantum). We will argue that the familiar structures of classical and quantum mechanics originate from OM.  In this sense, one may regard OM as a pre-quantum algebra.

\section{Expectation Values of the First Kind}
%operator expectation values can be generally formalized as linear functionals over Banach/Hilbert spaces with the structure of a $C^*$ algebra 

In the standard quantum-mechanical framework, expectation values of Hermitian operators can be found for a given quantum state in Hilbert space.    The expectation value refers to the probabilistic expected value of an experimentally measured quantity. For mixed states, one may construct the density matrix as a positive trace-class operator  $\rho = \sum_{i = 1}^n p_i |\psi_i\rangle \langle\psi_i|$.  The expectation value of self-adjoint operators ${\cal O}$ associated to observables is given by $Trace \left( \rho {\cal O} \right)$. Note that $\rho$ itself is defined in the Schr\"{o}dinger picture and requires the a priori specification of a complete set of basis states in the Hilbert space upon which the action of operators ${\cal O}$ is defined. On the other hand, if one starts with a pre-quantum operator algebra devoid of a Hilbert space as discussed in Sections 2 and 3,  how should one construct a density operator? Also, how does one interpret the trace of a given operator weighted by such a   density operator in the absence of a state space?

Let us proceed to define additional constructions within our OM framework that will enable us to answer the questions raised above. Within our operator algebra let us consider two sub-spaces $LinOp (X)$ and $LinOp (P)$ whose elements are linear combinations of operators $X$ and $P$ respectively with scalar multiplication given by a ring ${\cal R}$.
\begin{definition} 
We define maps 
\begin{eqnarray}
\phi(X) : X \to LinOp (X)  \qquad   \chi(P) : P \to LinOp (P)
\end{eqnarray}
where $\phi(X)$ and $\chi(P)$ respectively takes values in an ${\cal R}$-module whose basis are the generators of ${\cal L}$. 
\end{definition}
Here, we will consider the field of scalars ${\cal R}$ to be ${\mathbb C}$-valued (though that can be generalized to noncommutative scalars in an even more general setting, for instance, when one wants to discuss spinor-valued operators). 

Furthermore, the symplectic structure in OM endows the above maps with the following property: 
\begin{lemma}
For every $\phi(X)$, there exists an operator $\chi(P)$, such that 
\begin{eqnarray}
\frac{1}{\kappa }  \, {{\bf{Pr}}_m}  \, [ \phi(X), \, \chi(P) ]    \in {\mathbb R}_{\geq 0}  <  \infty 
\label{nm}
\end{eqnarray}
where ${{\bf Pr}_m}$ denotes an eigenvalue projection operator whose action on any $F \in {\cal L}$ yields the $m^{th}$ eigenvalue of $F$, that is, 
\begin{eqnarray}
{{\bf Pr}_m} \, F = f_m
\end{eqnarray}
for any given an $N$-dimensional representation of $F$ with eigenvalues $\{ f_m \}$ for $1 \leq m \leq N$. 
\end{lemma}
The proof of this lemma involves substituting $X^i \to P_i$ and $k \to k^{*}$ in $\phi(X) = \sum_{i = 1}^n k \, X^i$ for scalars $k \in {\mathbb C}$. 

Note that given an $N$-dimensional identity operator, we have ${{\bf Pr}_m} \,  {\mathbf 1}_N = 1$. Alternatively, instead of the  eigenvalue projection operator  ${{\bf Pr}_m}$, we could have also implemented the standard operator trace in Eq.~(\ref{nm}) above, which  would simply sum over all eigenvalues rather than projecting onto a specific one. That would also work for defining the density operator below. The operation  ${{\bf Pr}_m}$ is more useful when working with given operators $X^i$ and $P_i$ as we do here. On the other hand, the trace operation can be useful when considering an ensemble of $X^i$ and $P_i$ and performing a matrix model analysis.  

\begin{remark}
From Eq.~(\ref{nm}) we see that for given $\phi(X)$, a map $\chi(P)$ can be constructed in such a way that the above lemma holds. $\chi(P)$ thus constructed  can be thought of as the involution of $\phi(X)$. We will  denote this as $\overline{\phi(X)}$.  Eq.~(\ref{nm}) then defines a norm on $LinOp (X)$. 
\end{remark}
The right-hand side of Eq.~(\ref{nm})  can be verified using the canonical commutation relations.  

\begin{remark}
Furthermore,  $LinOp (X)$ almost carries the structure of a Hilbert space, except that the $\phi(X)$ are operators in an ${\cal R}$-module and we have not imposed completeness under the operator topology.  The space  $LinOp (X)$  of operators  (along with its involution operation and norm) is in fact  a  `{\it pre-Hilbert}' space ${\mathcal H}_{pre}$  (one would still need to show completeness under the operator topology in order for this to be a full-fledged  Hilbert space).  
\end{remark}

Equipped with  ${\mathcal H}_{pre}$  we can now define a new density operator $\hat{\rho}$ in OM: 
\begin{definition}  
 \begin{eqnarray}
\hat{\rho}  =  \frac{1}{\kappa^2 } \sum_i^n  w_i  \, | \, P_i \, ] \, [ \, X^i \, |
\label{rho}
\end{eqnarray}
which is expressed in the basis of ${\mathcal H}_{pre}$. The weights $w_i$ are real coefficients satisfying $w_i \geq 0$ and $\sum_{i = 1}^n w_i = 1$. 
\end{definition}  
The notation $[ \, X^i \, |$ indicates a basis operator with left-action on any operator $F$ such that taking the trace over ${\mathcal H}_{pre}$ results in the commutator  $[ \, X^i, \, F \, P_i \, ]$. Likewise, $| \, P_i \, ]$ denotes the conjugate operator, which due to the involution map, can also be expressed as $| \, \overline{X^i} \, ]$. Notice, that these are in fact the operator analogues of Dirac's bra-ket vectors, defined  directly using the canonical commutation relations of the Lie algebra ${\cal L}$. For that reason,  $[ \, X^i, \, F \, P_i \, ]$ is not a complex amplitude, but is  operator-valued in ${\cal L}$. 

Given the above definition of $\hat{\rho}$,  we then define operator expectation values as follows: 
\begin{definition}  
Given an operator $F$, we define its `expectation value of the first kind' as 
\begin{equation}
\angled{F}_m =   {{\bf Pr}_m} \, Tr_{{\mathcal H}_{pre}} (\hat{\rho} F)
\label{expval}
\end{equation}
where, we first take the trace over the bases of ${\mathcal H}_{pre}$, that is $[ \, X^i, \, \hat{\rho} F \, \overline{X^i} \, ]$, and then, apply the eigenvalue projection operator.  The expectation value so defined depends on the eigenvalue index $m$.  
\end{definition}  

\begin{remark}
The density operator $\hat{\rho}$ defined in Eq.~(\ref{rho})  satisfies:
\begin{equation}
Tr_{{\mathcal H}_{pre}} (\hat{\rho})  = {\mathbf 1}_N
\end{equation}
Moreover, $\hat{\rho}$ is also positive semi-definite and self-adjoint (where the adjoint operation is defined using the involution map on ${\mathcal H}_{pre}$ as conjugation, and the transpose is realized via flipping the operator bra and ket in $\hat{\rho}$), which justifies its use as a density operator for computing expectation values as in Eq.~(\ref{expval}). 
\end{remark}
For Hermitian operators  $F \in {\cal L}$ (that is, those with real eigenvalues), $\angled{F}_m$ will yield real values (for each $m$) that reflect the spectrum of $F$. The above constructions make use of the symplectic structure in OM and do not require any a priori notion of a Hilbert space of state vectors upon which   operators act. 
% potential theorem to prove about the 2 kinds of self-adjointedness and that the exp val are real - useful for a follow up 

Let us now examine the expectation value of the first kind for the operator $X^i$. Consider any given $N$-dimensional representation of $X^i$ with spectrum $\{ \lambda_m^i \}$ for $1 \leq m \leq N$. This may be taken to be an $N \times N$ matrix. Furthermore, the $X^i$ are Hermitian (as required for a Hermitian Hamiltonian). Therefore, the $\lambda_m^i$ are all real. The expectation value of $X^i$ then is:
\begin{equation}
\angled{X^i}_m =  {{\bf Pr}_m} \, Tr_{{\mathcal H}_{pre}} (\hat{\rho} X^i) = w_i \, \lambda_m^i 
\label{xval}
\end{equation}
which is now a function of $m$, the eigenvalue index (note that no summation over common indices is implied above).  Unlike the usual notion of an expectation value, which depends on the states of a system, $\angled{X^i}_m$ is a distribution depending on the eigenvalue index of the operator.  It will be instructive to now consider the continuum limit of the operator spectrum, where $\angled{X^i}_m$ for each given instance of $m$ can be thought of as defining a coordinate of a classical configuration space. To take the continuum limit of the eigenvalue spectrum of this operator we will use the same method typically employed in Hermitian matrix models (see \cite{brezin1978planar} or \cite{anninos2020notes} for a more recent survey). This limit  can be obtained by taking the large $N$ limit of the operator $X^i$ (using the representation of  $N \times N$ Hermitian matrices).  Consider  functions $\lambda^i (x)$ such that:
\begin{equation}
 \lambda^i (m/N) = \frac{\lambda_m^i}{N}
\label{nlimit}
\end{equation}
with the eigenvalues $\lambda_m^i$ being arranged in non-decreasing order $\lambda_1^i \leq \lambda_2^i \leq \cdots \leq \lambda_N^i$. Then in the large $N$ limit, we get $- \alpha \leq \lambda^i (x) \leq \alpha$ for some positive  $\alpha$ and $0 \leq x \leq 1$. The precise value of $\alpha$ does not concern us here, but it can be determined based on how the distribution of eigenvalues of these operators are constrained. Alternatively, one may consider an ensemble of random Hermitian matrices with an appropriate path integral and determine $\alpha$ from the resolvent of the corresponding loop equations as in \cite{brezin1978planar}. When $N \to \infty$ the scaling of eigenvalues in Eq.~(\ref{nlimit}) above yields a continuous spectrum over a finite support as shown in \cite{brezin1978planar}. 

\begin{remark}
The main reason for considering the aforementioned large $N$ limit of the operator expectation values (of the first kind) of  $X^i$ (and likewise $P_i$) is that this procedure  allows us to define new continuum variables   in terms of the spectrum of the operator $X^i$ (likewise for operators  $P_i$). In the following sections, we show that the new variables  
\begin{equation}
{\cal X}^i (x) \equiv \lim_{N \to \infty} \,  \frac{1}{N} \angled{X^i}_x \qquad  \qquad   {\cal P}_i (x) \equiv \lim_{N \to \infty} \,  \frac{1}{N} \angled{P_i}_x
\end{equation}
turn out to be a natural choice as coordinates of phase space. Here ${\cal X}^i$ and ${\cal P}_i$ are both continuous variables taking values over a subset of ${\mathbb R}^{2n}$. 
\end{remark}
Alternatively, in a matrix model analysis, the above phase space variables can also be obtained from the eigenvalue distribution of the matrix ensemble, in the large $N$ limit. 

Furthermore, below in Section 8, we have provided an additional perspective (not completely unrelated to the operator eigenvalue description above) whereby the set of points (${\cal X}^i$,  ${\cal P}_i$) acquire a spatial structure with a well-defined metric. This is achieved via entanglements of modes of operators $X^i$ and $P_i$, and is computed using the entanglement entropy based on a reduced density operator extracted from $\hat{\rho}$ defined above. In other words, that demonstrates how space emerges in OM from entanglements of operator  eigenmodes in the pre-Hilbert space ${\cal H}_{pre}$.

\section{Wave Functions, Observables and Expectation Values of the Second Kind}

The operator expectation values of the first kind defined above enables  construction of new variables $\angled{X^i}_m$ and $\angled{P_i}_m$, which in the continuum limit, yield the coordinates that will become relevant for classical phase space. In this sense, classical phase space is not a pre-existing fundamental space, but one that results from the pre-Hilbert structure within the OM framework. 
\begin{remark}
Algebraically, the expectation value of the first kind is a mapping from the noncommutative operator algebra associated to  ${\cal L}$  to a commutative algebra. This can be described in terms of  the following commutative diagram\footnote{For convenience of notation, we will drop the subscript $m$ in front of expectation values for the rest of this article, since the role of $\angled{\cdots}$ as a distribution is now clear. }: 
\begin{eqnarray}
\begin{tikzcd}[column sep=5em, row sep=5em]  
X^i   \arrow[d, "\frac{\partial }{\partial X^j}" '  ]     \arrow[r, " ",  shift left=0 ]  &  \angled{X^i}   \arrow[d,  "\frac{\partial }{\partial  \angled{X^j} }" ]      \\ 
   { [X^i,  P_j] }     \arrow[r, " ",  shift left=0 ]  &   \angled{ \, [X^i,  P_j] \, }
\end{tikzcd} 
\label{comdia1}
\end{eqnarray}
\end{remark}
If we require the commutative algebra on the right-hand side of the above diagram to preserve a Lie algebraic structure, then we will require that 
\begin{eqnarray}
 \angled{ \frac{1}{\kappa} \, [X^i,  P_j] \, }  &\cong&  \big \{ \angled{X^i},  \,  \angled{P_j}  \big \}_{PB} 
\label{compat1}
\end{eqnarray}
since the Poisson bracket is a natural choice for a Lie bracket associated to a commutative algebra. Eq.~(\ref{compat1}) is a Lie algebra preserving compatibility condition when going from a noncommutative Poisson structure to a commutative Poisson structure. The diagram in Eq.~(\ref{comdia1}) along with the above compatibility condition is formally a `dequantization' map from a noncommutative ${\cal C}^*$ algebra to a commutative one\footnote{A different realization of dequantization appears in terms of the Wigner map used by Groenewold and Moyal in the `phase space formulation' of quantum mechanics  \cite{groenewold1946principles,moyal1949quantum}.}   \cite{woodhouse1997geometric,ashtekar1999geometrical,hawkins2022quantization}.   

\begin{remark}
We can extend the diagram in Eq.~(\ref{comdia1}) to a second commutative square which admits complex-valued functions $f \in {\cal F}$ on the space spanned by the $\angled{X^i}$ (and in general, also spanned by $\angled{P_i}$).  Partial derivatives of these functions are expressed as their Poisson bracket with $\angled{P_i}$:  
\begin{eqnarray}
\begin{tikzcd}[column sep=5em, row sep=5em]  
\angled{X^i}  \arrow[d, "\frac{\partial }{\partial \angled{X^j} }" '  ]     \arrow[r, " ",  shift left=0 ]  &  f  \left(\angled{X^i}\right)   \arrow[d,  "\frac{\partial }{\partial  \angled{X^j} }" ]      \\ 
   \big \{ \angled{X^i},  \,  \angled{P_j}  \big \}_{PB}     \arrow[r, " ",  shift left=0 ]  &   \big \{ f \left(\angled{X^i}\right),  \,  \angled{P_j}  \big \}_{PB}  
\end{tikzcd} 
\label{comdia2}
\end{eqnarray}
\end{remark}
Admitting an algebra of functions on the space spanned by $\angled{X^i}$ and $\angled{P_i}$  yields a bundle over ($\angled{X^i}$, $\angled{P_i}$), with $\frac{\partial }{\partial  \angled{X^i} }$ and $\frac{\partial }{\partial  \angled{P_i} }$ as tangent vectors. We will call this the phase space bundle ${\cal E}$ over the base over ($\angled{X^i}$, $\angled{P_i}$). The aforementioned functions can be thought of as sections in ${\cal E}$. Furthermore, working in the large $N$ limit, we  replace $\angled{X^i} \to {\cal X}^i$ and $\angled{P_i} \to {\cal P}_i$. Within this continuum limit, both  classical and quantum wave functions $\psi_{KvN}$ and  $\psi_{QM}$, respectively,  are sections in 
 ${\cal E}$.  The difference between these sections is related to quantization (discussed in Section 7).

Hence, wave functions are complex-valued constructs existing in a phase space bundle ${\cal E}$.  There are two kinds of wave functions alluded to above are: (i) the classical Koopman-von Neumann wave function $\psi_{KvN} \in {\cal F}  \left( {\cal X}, {\cal P} \right)$, which depends on both ${\cal X}$ and ${\cal P}$;  and (ii) the standard quantum mechanical wave function  $\psi_{QM} \in {\cal F}  \left( {\cal X} \right)$, which depends only on $\angled{X}$. 

\begin{remark}
The Koopman-von Neumann (KvN)  formulation of  mechanics was originally developed in the 1930s by Koopman and von Neumann as a theory of classical ensembles, with a particular focus towards ergodic theory  
 \cite{koopman1931hamiltonian,neumann1932operatorenmethode}.  It makes use of a similar technical machinery to quantum mechanics, namely, Hilbert spaces, wave functions as states and operators acting upon wave functions. For this reason,  Koopman-von Neumann mechanics has also been relevant to the conceptual foundations of quantum theory   \cite{bondar2012operational,wilczek2015notes,klein2018koopman}.  The wave function $\psi_{KvN}$ in Koopman-von Neumann mechanics is a complex-valued function defined using the classical  phase space distribution function $\tilde{\rho} ({\cal X},  {\cal P}; t) = \psi_{KvN}^\ast \, \psi_{KvN}$, and $\psi_{KvN}$ depends on both  ${\cal X}$ and  ${\cal P}$. Since $\tilde{\rho} ({\cal X},  {\cal P}; t)$ is a probability density, this implies a normalization condition on $\psi_{KvN}$. The latter induces an inner product on the space of Koopman-von Neumann wave functions (see Eq.~(\ref{ncond})),  thus leading to a Hilbert space of states ${\cal H}_{KvN}$.   The Koopman-von Neumann equation is the classical analogue of the Schr\"{o}dinger's equation and follows from Liouville's theorem  (this is derived in Eq.~(\ref{kvn}) below).  
\end{remark}

Now let us investigate the algebra of operators acting on  functions in ${\cal E}$.  These are $\angled{X^i}$ and $\angled{P_i}$, which act by ordinary multiplication; and the derivatives indicated in Eq.~(\ref{comdia2}). The latter can equivalently be expressed as the following Poisson brackets:   
 $\big \{ \bullet,  \,  \angled{P_i}  \big \}_{PB}$ and  $\big \{ \angled{X^i},  \,  \bullet  \big \}_{PB}$.  One easily checks that these operators satisfy the following commutation relations:
\begin{eqnarray}
[{\cal X}^i,  \, {\cal P}_j] = 0  \qquad  \qquad   [{\cal X}^i,  \, \tilde{\kappa}  \big \{  {\cal P}_j,  \,  \bullet  \big \}_{PB}] &=& \tilde{\kappa} \, \delta^i_j  \nonumber  \\
\, [ \tilde{\kappa} \big \{  {\cal X}^i,  \,  \bullet  \big \}_{PB},  \, \tilde{\kappa} \big \{  {\cal P}_j,  \,  \bullet  \big \}_{PB}  ]  =  0   \qquad  \qquad     [ \tilde{\kappa} \big \{  {\cal X}^i,  \,  \bullet  \big \}_{PB},  \,   {\cal P}_j ] &=&  \tilde{\kappa}  \, \delta^i_j   \nonumber  \\   
\!  [{\cal X}^i,  \, \tilde{\kappa}  \big \{  {\cal X}^j,  \,  \bullet  \big \}_{PB}] =  0   \, \qquad  \qquad     [  {\cal P}_i ,  \, \tilde{\kappa}  \big \{  {\cal P}_j,  \,  \bullet  \big \}_{PB}  ]   &=&   0
\label{calops}
\end{eqnarray}
where we will be working with these operators in the continuum limit of their respective  variables. The constant $\tilde{\kappa}$ has been introduced here to match dimensions. Note that $\tilde{\kappa}$ need not be the same as the $\kappa$ in OM. The operators in Eq.~(\ref{calops}) act on wave functions in ${\cal E}$,   whereas those in the OM canonical commutation relations  are defined without any reference to a space of states. When we arrive at the Schr\"{o}dinger's equation,  $\tilde{\kappa}$ will be related to $i \, \hbar$, whereas $\kappa$ is still a free parameter that may depend on another fundamental scale. 

From the perspective of ${\cal E}$, the above operator algebra in Eq.~(\ref{calops})  and the associated commutation relations  hold for both, classical Koopman-von Neumann mechanics as well as traditional quantum mechanics. The difference between the two will be identified as a "twisting" of sections following a different identification of momentum in the two cases. 

Being ${\mathbb C}$-valued functions, both  $\psi_{KvN}$ and $\psi_{QM}$  admit the  $L^2$-norm over phase space and configuration space respectively
\begin{eqnarray}
\int d {\cal X} d {\cal P} \, \psi_{KvN}^\ast \, \psi_{KvN}  = 1  \qquad \qquad  \int  d {\cal X}  \, \psi_{QM}^\ast \, \psi_{QM}  = 1
\label{ncond}
\end{eqnarray}
And it is precisely this norm that endows the space of wave functions with the structure of a Hilbert space. In other words, this is a Hilbert space of sections over the phase space bundle ${\cal E}$. We will denote this Hilbert space as ${\cal H}_{KvN/QM}$. It is this $L^2$-norm in ${\cal H}_{KvN/QM}$ with respect to which the familiar density operator for a system in a mixed state is defined as: 
\begin{eqnarray}
\rho  =   \sum_j  p_j  \, | \, {\psi_j}_{KvN/QM}  \, > \, < \, {\psi_j}_{KvN/QM}  \, |
\end{eqnarray}
where each of the pure states ${\psi_i}_{KvN/QM}$ occurs with probability $p_j$. Using this, expectation values of observables $F$ are then defined as
\begin{eqnarray}
\angled{F} = tr \left( \rho \, F  \right)
\end{eqnarray}
We will call these expectation values of the second kind to distinguish them from the pre-quantum expectation values of the first kind. The above Hilbert space and expectation values exist for both, Koopman-von Neumann classical mechanics as well as standard quantum mechanics. On this Hilbert space, an   expectation value (of the second kind) corresponding to an observable is a  function over states. But this object also has a geometric interpretation in ${\cal E}$, where it quantifies the extent to which an  operator $\hat{A}$ deforms a section $\psi_{KvN/QM}$ over ${\cal E}$ (with undeformed sections yielding the identity).  

Hence, expectation values of the second kind are simply the standard expectation values with respect to a Hilbert space wave function that appear in 
Koopman-von Neumann mechanics as well as quantum mechanics. 
%Geometrically, this relates points and vectors in ${\cal E}$ to functions on ${\cal H}_{KvN/QM}$. 
\begin{remark}
The map from expectation values of the first kind to those of the second kind lead to the  following commutative diagram:
\begin{eqnarray}
\begin{tikzcd}[column sep=5em, row sep=5em]  
\angled{X^i}  \arrow[d, "\frac{\partial }{\partial \angled{X^j} }" '  ]     \arrow[r, " ",  shift left=0 ]  & \angled{ \angled{X^i} }    \arrow[d,  "\frac{\partial }{\partial  \angled{ \angled{X^j} } }" ]      \\ 
   \big \{ \angled{X^i},  \,  \angled{P_j}  \big \}_{PB}     \arrow[r, " ",  shift left=0 ]  &   \big \{ \angled{\angled{X^i}},  \,  \angled{\angled{P_j}}  \big \}_{PB}  
\end{tikzcd} 
\label{comdia3}
\end{eqnarray}
\end{remark}
This diagram represents the transition from "points" in ${\cal E}$ to ensemble averages.  

All the above three commutative diagrams realize algebraic homomorphisms   from ${\cal L}$ to  ${\cal E}$ to ${\cal H}_{KvN/QM}$. In principle, these maps can also be formalized using functorial constructions from the appropriate  categories of noncommuting operators to those of commuting entities. In particular, ${\cal F} \left(\angled{X^i}\right)$ would be replaced by the relevant category of Hilbert spaces. The  categorification of the commutative diagrams above would presumably encapsulate the formalism of categorical quantum mechanics as a special case \cite{abramsky2009categorical}. 

The maps constructed above also suggests an interesting generalization of the wave function itself. Instead of considering only the algebra of functions over expectation values of $X^i$, one may investigate the algebra of functions over expectation values of operator products, in other words, functions of n-point operator products. Given that the $\angled{X^i}$ have an interpretation as points of space, higher products refer to $n$-point geometric correlations such as extended objects (geodesics, membranes, etc)\footnote{The proper formalism for computing such quantities will require a field theoretic extension of what we have done in this work and might have possible connections to the formalism of meta-string theory developed in  \cite{freidel2015metastring,freidel2016quantum}.}.

\section{Pre-Quantum Origins of Ehrenfest's Theorem }

Before we discuss quantization and Schr\"{o}dinger's equation in the next section, let us discuss one more result whose generalization turns out to be relevant to pre-quantum formalizations: Ehrenfest's theorem.  The standard   version of Ehrenfest's theorem is usually stated as a prescription to realize classical-like laws for expectation values of quantum mechanical observables. Here, we demonstrate  that there are in fact multiple manifestations of Ehrenfest's theorem, or what could be called Ehrenfest-like theorems, obtained by applying  expectation values of the first and second kind, starting from the OM pre-quantum algebra. 

Firstly, recall that the Hamilton's equations in operator form already show up at the level of the OM algebra, without recourse to  classical calculus.  
Furthermore, for a Hamiltonian quadratic in momentum, this yields: 
\begin{eqnarray}
\frac{d X^i}{d t}  &=&  \frac{1}{m}  P_i   \label{eq37}  \\
\frac{d P_i}{d t}  &=&  -  \frac{\partial V}{\partial X^i}    \label{eq38}   
\end{eqnarray} 
which is simply the first-order representation of Newton's laws for operators $X^i$ and $P_i$.  What we demonstrate below is that all manifestations of Ehrenfest's theorem originate from these operator relations by applying the algebraic homomorphisms, discussed in the previous section, from the noncommutative algebra to the commutative one. 

Now, let us apply the expectation value map of the first kind to the operatorial  Hamilton's equations above. This will map (alternatively, one may think of this as a functorial lift) these equations to dynamics in the regime of commutative algebras, otherwise known as classical mechanics. To see this, let us use the  following generic form of the non-relativistic Hamiltonian operator that is  quadratic in momentum and carries an arbitrary potential: 
\begin{eqnarray}
H = \frac{1}{2 m}  \sum_{i,j = 1}^n g^{ij} P_i P_j + V (\{ X^i \})
\label{H}
\end{eqnarray}
where the inner product $g^{ij}$ will refer to the flat metric $\eta^{i j}$ for what follows (and is not operator valued here). Using this Hamiltonian in the OM equation of motion, Eq.~(\ref{hameq2}), and applying the operator expectation of the first kind on both sides of this equation yields: 
\begin{eqnarray}
\kappa \, \Big \langle \frac{d X^i}{d t} \Big \rangle &=& \big \langle  [X^i,  H] \big \rangle  \, = \,  \kappa \, \frac{\eta^{i j}}{m} \langle P_j  \rangle
\label{24}
\end{eqnarray}
While the right-hand side is proportional to $\langle P_j  \rangle$, we now want to understand how the left-hand side of this equation relates to $\langle X^i  \rangle$.  To see this, we first need a definition for time derivatives of    commuting quantities $f$, which are functions of continuous variables ${\cal X}$ and ${\cal P}$ (working in the large $N$ limit will ensure continuity of our phase space variables).  

In the noncommutative OM calculus, the time derivative was defined as a Lie bracket of noncommutative operators. However, in order work with a commutative calculus that carries an analogous Lie-algebraic structure, we need to define the following commutative time derivative:
\begin{definition} 
The total time derivative of ${\mathbb C}$-valued  functions of variables  ${\cal X}$, ${\cal P}$ and time is defined via a  Poisson bracket involving a Hamilton function in the variables ${\cal X}$ and ${\cal P}$ such that: 
\begin{eqnarray}
 \frac{d \, f  }{d t_{c}}  &=&  \frac{\partial \, f  }{\partial \, t_{c}}  \, + \,  \big \{ f,  \,  H_{cl} \big \}_{PB}  
\label{ex3}
\end{eqnarray} 
where the partial derivative  $\partial / \partial t_{c}$ is the same as partial derivatives in standard analysis.

The Hamilton function $H_{cl}$ above, refers to a classical system, expressed in terms of continuous  commuting  phase space variables  ${\cal X}^i$ and ${\cal P}_i$,  given by: 
\begin{eqnarray}
H_{cl} \, = \,  \frac{1}{2 \, m}  \sum_{i = 1}^n  {\cal P}_i^2  \,  + \, V \left( \{ {\cal X}^i \} \right)
\label{ex4}
\end{eqnarray}
and the classical Poisson bracket is expressed in terms of classical partial derivatives:  
\begin{eqnarray} 
\big \{ f,  \, g  \big \}_{PB}   &=&  \sum_{i = 1}^n  \left(  \frac{\partial \, f }{\partial \, {\cal X}^i}  \frac{\partial \, g }{\partial \, {\cal P}_i}  \, - \,  \frac{\partial \, f }{\partial \, {\cal P}_i}  \frac{\partial \, g }{\partial \, {\cal X}^i}   \right) 
\label{ex5}
\end{eqnarray}
\end{definition}
Here we will use the notation $d / dt_{c}$ and $\partial / \partial t_{c}$ for  commutative time derivatives   (as opposed to $d / dt$, which denotes the time derivative defined for a noncommutative calculus referring to derivatives of operators). This notation is simply meant to distinguish between different types of derivatives associated to the algebra that they act upon.  The definition in Eq.~(\ref{ex3})  is  a  Lie algebra-preserving homomorphism from ${\cal L}$ to a commutative algebra, where the expectation value of the first kind maps operators in OM to functions $\angled{\cdots}: F \to f$.  The partial derivative   $\partial / \partial t_{c}$ in the above definition has been included to account for functions $f$ with explicit time dependence. 

Note that instead of proposing Eq.~(\ref{ex3})  as a definition, one could have also derived it from classical calculus. However, the derivation would involve assuming Hamilton's equations in order to get  Poisson brackets on the right-hand side. The advantage of taking  Eq.~(\ref{ex3})  as a definition is that  Hamilton's equations  then follow from this commutative algebra. To see this one can simply substitute ${\cal X}^i$ respectively ${\cal P}_i$ into Eq.~(\ref{ex3}) to get precisely the two Hamilton's equations in classical mechanics.  

Now returning to the Ehrenfest's theorem.  Given the above definition of commutative derivatives, the time derivative of ${\cal X}^i$ is obtained as:  
\begin{eqnarray}
 \frac{d \, {\cal X}^i }{d t_{c}}    \, = \,  \frac{1}{m}  {\cal P}_i  \, = \,  \frac{1}{m}   \frac{1}{N} \angled{P_i} 
\label{ex6}
\end{eqnarray}
where the first equality is just Newton's law for the commutative variable ${\cal X}^i$ (for a Hamiltonian quadratic in momentum). Compare this to Eq.~(\ref{eq37}). The second equality in Eq.~(\ref{ex6})  holds only in the large $N$ limit. In that limit the right-hand side of Eq.~(\ref{ex6}) is comparable to the right-hand side of Eq.~(\ref{24}). This  implies  that the expectation value of the operator time derivative of $X^i$  can be replaced by the commutative  time derivative of the expectation value of  $X^i$ in the continuum limit. That is:  
\begin{eqnarray}
\frac{1}{N}  \Big \langle \frac{d X^i}{d t} \Big \rangle  \,  \to \,  \frac{d \, {\cal X}^i }{d t_{c}}   
\label{ex1}
\end{eqnarray}
in the large $N$ limit. 
%which maps the time derivative of $X^i$ in OM, using the expectation value map, to the corresponding time derivative in a commutative algebra involving continuous variables.   
%This is one way to express Ehrenfest's theorem for the operator $X^i$, but now derived using the operator expectation value Eq.~(\ref{expval}) and operator time derivative Eq.~(\ref{dt}), defined in OM. 
Now  implementing this mapping in Eq.~(\ref{24}) and expressing everything in  continuum variables   ${\cal X}^i$ and ${\cal P}_i$  gives Newton's law for the commutative variable ${\cal X}^i$ (for a Hamiltonian quadratic in momentum) from the OM expectation value equation (Eq.~(\ref{24})).  This serves as a consistency check for the roles of ${\cal X}^i$ and ${\cal P}_i$ as classical phase space variables.

In a similar manner, an Ehrenfest-like theorem for the time derivative of the momentum operator in OM is obtained by inserting the Hamiltonian operator above into Eq.~(\ref{hameq1}) and taking the operator expectation values of the first kind on of both sides to get: 
\begin{eqnarray}
\kappa \, \Big \langle \frac{d P_i}{d t} \Big \rangle &=& - \, \kappa \, \Big \langle \frac{\partial \, V( \{ X^i \} ) }{\partial X^i} \Big \rangle 
\label{ehr2}
\end{eqnarray}
This is the statement of an Ehrenfest-like theorem for the momentum operator in OM. Given that the expectation value map of the first kind relates to classical phase space variables, the above statement can be cast in the traditional form of Newton's law for quadratic potentials. This can be seen as follows.  
Consider the special case when the potential is of the form  $V (  \{ X^i \} ) \, = \, g  \, \delta_{i_1 i_2 \cdots i_l} \, X^{i_1} X^{i_2} \cdots X^{i_l}$ with coupling constant $g$. We then have:
\begin{eqnarray}
 \Big \langle \frac{d P_i}{d t} \Big \rangle &=& - \, g \,  l  \, \delta_{i j \cdots j}  \,  \big \langle \left( X^j \right)^{ l - 1 }  \big \rangle
\label{31}
\end{eqnarray} 
On the other hand, operating the commutative time derivative on ${\cal P}_i$   yields:  
\begin{eqnarray}
 \frac{d \, {\cal P}_i }{d t_{c}}  &=&  - \,  \frac{\partial \, V \left( \{ {\cal X}^j  \} \right) }{\partial \, {\cal X}^i }  \, =  \,  - g  \,  l  \, \delta_{i j \cdots j}  \,  \left( {\cal X}^j  \right)^{l - 1}  \, =  \,  - g  \,  l  \, \delta_{i j \cdots j}  \,  \frac{1}{N^{l - 1}} \big \langle  X^j  \big \rangle^{l - 1} 
\end{eqnarray}
where the last equality holds in the large $N$ limit. 

Now, when $l = 2$, we have:
\begin{eqnarray}
\frac{1}{N}  \Big \langle \frac{d P_i}{d t} \Big \rangle  \,  \to \,  \frac{d \, {\cal P}_i }{d t_{c}}   
\label{ex2}
\end{eqnarray}
in the large $N$ limit; thus replacing the expectation value of the OM time derivative of $P_i$  with the  commutative time derivative of the expectation value of  $P_i$ in the continuum limit. For other values of $l$, this relation only holds modulo statistical (higher-order) corrections. Hence, implementing Eq.~(\ref{ex2}) in Eq.~(\ref{ehr2}) with the above-mentioned polynomial potential yields Newton's law for the commutative variable ${\cal P}_i$ upon applying expectation values of the first kind to the OM equation of motion and taking the large $N$ limit (when $l = 2$ in the polynomial potential).

Eqs.~(\ref{24}) and (\ref{ehr2}) are Ehrenfest-like  theorems  with respect to expectation values of the first kind in OM. In the large $N$ limit, these equations  are equivalent to standard Newton's laws in the case of a quadratic Hamiltonian. 

On the other hand, the origins of the classical Hamilton's equations can be directly traced to the Lie algebra-preserving homomorphism from OM to the commutative setting where operators map to their expectation values of the first kind and Lie brackets map to Poisson brackets.

%Moreover, this procedure for obtaining time derivatives of commutative functions starting from operators  will be important for deriving  Schr\"{o}dinger's equation in OM, once we make a suitable ansatz for an operator that serves as the precursor to the traditional  quantum mechanical wave function.  

Furthermore, subsequent homomorphisms from expectation values of the first kind to those of the second kind lead to other realizations of  Ehrenfest-like  theorems. We will now examine these. Let us first consider the case for Koopman-von Neumann mechanics, where expectation values depend on wave functions in the Hilbert space ${\cal H}_{KvN}$. There are two ways to apply the two kinds of expectation values. The first way involves starting with the equation of motion in  OM and applying successive expectation values of the first and  second kind to get:
\begin{eqnarray}
 \angled{\angled{  \frac{d X^i}{d t} }}  &=&   \frac{\eta^{i j}}{m} \angled{\angled{P_j}} 
%\label{24}
\end{eqnarray}
which in the large $N$ limit, is equivalent to 
\begin{eqnarray}
 \frac{d \, \angled{{\cal X}^i} }{d t_{c}}    \, = \,  \frac{1}{m}  \angled{{\cal P}_i}  
%\label{ex6}
\end{eqnarray}
computed using Poisson brackets in term of variables $\angled{\angled{X^i}}$ and $\angled{\angled{P_j}}$; and the Hamilton function of an "average" particle from a classical ensemble:
\begin{eqnarray}
H_{avg} \, = \,  \frac{1}{2 \, m}  \sum_{i = 1}^n  \angled{{\cal P}_i}^2  \,  + \, V \left( \{ \angled{{\cal X}^i} \} \right)
%\label{ex4}
\end{eqnarray}
where the $\angled{\angled{X^i}}$ and $\angled{\angled{P_j}}$ are expectation values with respect to the Koopman-von Neumann wave function $\psi_{KvN}$. 

Likewise, for the time derivative of the momentum operator, we have the following Ehrenfest-like theorem
\begin{eqnarray}
\angled{ \angled{ \frac{d P_i}{d t}  } } &=&  -   \,  \angled{ \angled{ \frac{\partial \, V( \{ X^i \} ) }{\partial X^i}  }  } 
\end{eqnarray}
which in the special case of a fully quadratic Hamiltonian is equivalent to (in the large $N$ limit):
\begin{eqnarray}
 \frac{d \, \angled{{\cal P}_i} }{d t_{c}}    \, = \,  -   \frac{\partial \,  V \left( \{ \angled{{\cal X}^i} \} \right) }{\partial \, \angled{{\cal X}^i} } 
\end{eqnarray}
for an "average" particle from a classical ensemble. 

The other way to apply the two kinds of expectation values would be to start with ${\cal X}$ and ${\cal P}$, which are already expectation values of the first kind, and apply expectation values of the second kind to their time derivatives (with respect to the wave function  $\psi_{KvN}$). In Koopman-von Neumann mechanics ${\cal X}$ and ${\cal P}$ also play the role of classical operators, whose time derivatives are defined via the Poisson bracket. With this prescription, we get:
\begin{eqnarray}
 \angled{ \frac{d \,  {\cal X}^i }{d t_{c}}  }  \, = \,  \frac{1}{m}  \angled{{\cal P}_i}    \qquad \qquad    \angled{  \frac{d \,  {\cal P}_i }{d t_{c}}  }   \, = \,  -  \angled{ \frac{\partial \,  V \left( \{  {\cal X}^i \} \right) }{\partial \, {\cal X}^i }  }
\label{kneh}
\end{eqnarray}
Furthermore, notice that in the so-called Schr\"{o}dinger picture, operators carry a time dependence, but states do not. We can therefore pull the time derivatives out of the expectation value operation (defined in terms of $\psi_{KvN}$) to get:
\begin{eqnarray}
 \angled{ \frac{d \,  {\cal X}^i }{d t_{c}}  }  \, = \,  \frac{d \,  \angled{{\cal X}^i} }{d t_{c}}   \qquad \qquad    \angled{  \frac{d \,  {\cal P}_i }{d t_{c}}  }   \, = \,  \frac{d \,  \angled{{\cal P}_i} }{d t_{c}}
\end{eqnarray}
Substituting this into Eq.~(\ref{kneh}) above gives the standard Ehrenfest's theorem for Koopman-von Neumann mechanics. In the next section we show that similar considerations yield the standard form of Ehrenfest's theorem for the quantum mechanical wave function $\psi_{QM}$.

In summary, the Ehrenfest's theorem and its related manifestations all originate from the operator form of Hamilton's equations and successive Lie algebra preserving  homomorphisms. The insight that this exercise gives is that   Hamilton's equations are not merely classical laws, but universal laws of dynamics  capturing evolution of symplectic variables, which manifest at the level of the pre-quantum operator algebra, the classical commutative algebra (for individual particles), the classical ensemble algebra (for Koopman-von Neumann distributions), and the algebra of quantum mechanical observables (we shall see this case in the following section).  In fact, in  \cite{bondar2012operational}  it was even suggested that the Ehrenfest's theorem should be thought of as more fundamental than both, the Koopman-von Neumann's as well as the Schr\"{o}dinger's equation; and those authors seek to derive the latter two equations from a version of Ehrenfest's theorem. Here we will demonstrate a different route to obtaining Koopman-von Neumann's and Schr\"{o}dinger's equation, which nonetheless makes use of Hamilton's equations in one of its many guises.

\section{From Koopman-von Neumann's to Schr\"{o}dinger's Equation}
 
In Section 5, we identified the Koopman-von Neumann wave function $\psi_{KvN}$ as belonging to the algebra of  functions $f \left( {\cal X},  {\cal P} \right)$ over continuum variables ${\cal X}$ and ${\cal P}$, where the latter are   obtained as  expectation values of the first kind associated to operators $X$ respectively $P$ of the OM pre-quantum algebra\footnote{Recall that variables ${\cal X}$ and ${\cal P}$ without subscripts or superscripts are merely shorthand notation for the $n$-tuple of components $\{ {\cal X}^i \}$ and $\{ {\cal P}_i \}$.}.  In this section, we discuss how $\psi_{KvN}$ relates to the quantum wave function $\psi_{QM}$ and in the process we will recover  Schr\"{o}dinger's equation from the classical Koopman-von Neumann equation. 

Geometrically, $\psi_{KvN}$ forms a section of the phase space bundle ${\cal E}$, and the space of these sections along with the norm induced from $L^2$ functions, yields the Hilbert space ${\cal H}_{KvN}$. In that sense, the origins of  $\psi_{KvN}$ can be traced to the OM pre-Hilbert space ${\cal H}_{pre}$ of operators and successive homomorphisms bring us to the familiar space of wave functions ${\cal H}_{KvN}$. 

Analogously, the same is true for the wave function  $\psi_{QM}$ in quantum mechanics  and its familiar home in ${\cal H}_{QM}$, with the caveat that the former depends only on ${\cal X}$ and not on ${\cal P}$.  We will trace this subtle difference between  $\psi_{KvN}$ and $\psi_{QM}$ to a quantization scheme interpreted as a "twisting" of sections over  ${\cal E}$. We will first write down the Koopman-von Neumann equation for $\psi_{KvN}$ and then the Schr\"{o}dinger equation from it via quantization. 

Note that the Koopman-von Neumann equation follows from Liouville's theorem, which posits the constancy of the phase space distribution function $\tilde{\rho} ({\cal X},  {\cal P}; t)$ along any trajectory. Expressing $\tilde{\rho} ({\cal X},  {\cal P}; t)$ in terms of a complex-valued function  $\psi_{KvN}$ as $\tilde{\rho} ({\cal X},  {\cal P}; t) = \psi_{KvN}^\ast \, \psi_{KvN}$ yields the Koopman-von Neumann equation (and its conjugate). Hence, we have: 
\begin{eqnarray}
   \frac{\partial \, \psi_{KvN} ({\cal X},  {\cal P}) }{\partial t_{c}}  &=& - \sum_{i = 1}^n  \left( \frac{d {\cal X}^i }{d t_{c}} \,  \frac{\partial \, \psi_{KvN} }{\partial \, {\cal X}^i } \, + \,   \frac{d {\cal P}_i }{d t_{c}} \,  \frac{\partial \, \psi_{KvN} }{\partial \, {\cal P}_i }  \right)  
\end{eqnarray}
For the time derivatives of ${\cal X}^i$ and ${\cal P}_i$ we then substitute the version of  Hamilton's equations for a commutative algebra. This yields precisely the Poisson bracket in terms of the classical Hamilton function $H_{cl}$ on the right-hand side:
\begin{eqnarray}
   \frac{\partial \, \psi_{KvN}  }{\partial t_{c}}  &=&  \big \{  H_{cl},  \,  \psi_{KvN}  \big \}_{PB}
\label{kvn}
\end{eqnarray}
which is the familiar form of the Koopman-von Neumann equation, expressed in terms of the Liouville operator, acting on the classical wave function $\psi_{KvN}$.   The Koopman-von Neumann equation pertains to classical mechanics, albeit from a statistical perspective (with the classical Liouville equation expressing the statistical density function).  Notice the important role Hamilton's equations play here.  In our framework,  Hamilton's equations result  from mapping structures in ${\cal L}$ to the corresponding commutative setting, following which the wave function  $\psi_{KvN}$  originates from sections in ${\cal E}$.  

Of course, there are also  other ways to derive or generalize the Koopman-von Neumann equation (for instance, see  \cite{bondar2012operational,klein2018koopman}).  However, here our objective  is in investigating the underlying algebraic structures behind this equation and ultimately, its relation to the Schr\"{o}dinger's equation.   

We now show that a simple quantization map applied to the Koopman-von Neumann equation  leads to  Schr\"{o}dinger's equation. To see how this works, let us recall the operators identified earlier in  Section 5, which act on sections of the bundle ${\cal E}$. These are ${\cal X}^i$,  ${\cal P}_i$,  $\tilde{\kappa}  \big \{  {\cal X}^i,  \,  \bullet  \big \}_{PB}$ and  $\tilde{\kappa}  \big \{  {\cal P}_i,  \,  \bullet  \big \}_{PB}$. They satisfy the algebra given in Eq.~(\ref{calops}). Notice that this algebra contains remnants of both classical and quantum mechanics and is applicable to generic sections within the bundle ${\cal E}$. To understand what constitutes a classical versus a quantum wave function, the key issue is to identify the relevant observables. Given that the wave function represents the physical state of the system, it should only depend on quantities that are observable. The classical wave function $\psi_{KvN}$  only depends on ${\cal X}^i$ and ${\cal P}_i$ corresponding to the observable position and momentum operators. And given that these operators commute, they are simultaneously diagonalizable, resulting in joint eigenstates of ${\cal X}^i$ and ${\cal P}_i$ for the wave function $\psi_{KvN}$. 

The key observation that guides us in transitioning to the quantum wave function $\psi_{QM}$ is that the observable momentum operator from the algebra in Eq.~(\ref{calops}) is now $\tilde{\kappa}  \big \{  {\cal P}_i,  \,  \bullet  \big \}_{PB}$ and not ${\cal P}_i$. And given that ${\cal X}^i$  does not commute with the new momentum operator, these two observables are complementary. Hence,  $\psi_{QM}$ can be expressed either in the eigenbasis of ${\cal X}^i$ or $\tilde{\kappa}  \big \{  {\cal P}_i,  \,  \bullet  \big \}_{PB}$, with the former choice often referred to as a "projection to configuration space". From the above perspective, quantization is simply a mapping from one sub-algebra of Eq.~(\ref{calops}) to another:
\begin{eqnarray}
{\cal P}_i \to \frac{\tilde{\kappa}}{2}  \big \{  {\cal P}_i,  \,  \bullet  \big \}_{PB}   \qquad  \qquad  \tilde{\kappa}  \big \{  {\cal X}^i,  \,  \bullet  \big \}_{PB}  \to {\cal X}^i
\label{alghalf}
\end{eqnarray}
which from the point of view of sections of the bundle ${\cal E}$ can be geometrically interpreted as "twisting" a classical section that can be deformed in the directions of ${\cal X}^i$, ${\cal P}_i$,  $\tilde{\kappa}  \big \{  {\cal X}^i,  \,  \bullet  \big \}_{PB}$ and  $\tilde{\kappa}  \big \{  {\cal P}_i,  \,  \bullet  \big \}_{PB}$ to a quantum section that can only be deformed in the directions of ${\cal X}^i$ and  $\tilde{\kappa}  \big \{  {\cal P}_i,  \,  \bullet  \big \}_{PB}$ (which forces the projection of the wave function to configuration space). The quantized system does not depend on ${\cal P}_i$ anymore, and this is now not the physical momentum. Instead, $\tilde{\kappa}  \big \{  {\cal P}_i,  \,  \bullet  \big \}_{PB}$ is the new momentum (operator). 

Implementing the quantization scheme above into the Koopman-von Neumann equation (Eq.~(\ref{kvn})) directly leads to:
\begin{eqnarray}
   \frac{\partial \, \psi_{QM} ( {\cal X}, \tilde{\kappa} \big \{  {\cal P},  \,  \bullet  \big \}_{PB} ) }{\partial  t_{c}}  &=& - \sum_{i = 1}^n  \left( \frac{\tilde{\kappa}}{2 m} \big \{  {\cal P}_i,  \,  \bullet  \big \}_{PB}  \,  \frac{\partial \, \psi_{QM} }{\partial \, {\cal X}^i } \, - \,  \frac{\partial \, V ({\cal X}) }{\partial \, {\cal X}^i }  \,  \frac{ {\cal X}^i }{\tilde{\kappa}}  \,  \psi_{QM}   \right)  
\end{eqnarray}
which when $\tilde{\kappa} = i \, \hbar$, is exactly the Schr\"{o}dinger's equation:
\begin{eqnarray}
\tilde{\kappa}  \,   \frac{\partial \, \psi_{QM} }{\partial t_{c}}  &=& \sum_{i = 1}^n  \frac{\eta^{i j} \, \tilde{\kappa}^2 }{2 m}  \big \{  {\cal P}_i,  \,  \bullet  \big \}_{PB}  \, \big \{  {\cal P}_j,  \,  \bullet  \big \}_{PB}  \, \psi_{QM}  \, + \,  \widetilde{V} ({\cal X})  \, \psi_{QM}      
\end{eqnarray}
with a shifted potential 
\begin{eqnarray}
\widetilde{V} ({\cal X})   \, = \, - V ({\cal X})   \, + \, \sum_{i = 1}^n   \frac{\partial \, V ({\cal X}) \cdot {\cal X}^i }{\partial \, {\cal X}^i } 
\end{eqnarray}
For polynomial potentials, this shift amounts to a scaling of couplings of the potential. For instance, consider the case with $i = 1$. Then: 
\begin{eqnarray}
 \sum_{k = 1}^m  a_k \,  ( {\cal X}^1 )^k  \,  \to  \,  \sum_{k = 1}^m \,  k \, a_k \,  ( {\cal X}^1 )^k    
\end{eqnarray}
Hence, with the above quantization scheme, we get the following Hamiltonian operator for the quantized system:
\begin{eqnarray}
H_{qm} \, = \,  \frac{\eta^{i j}}{2 \, m}  \sum_{i = 1}^n   \tilde{\kappa}^2  \big \{  {\cal P}_i,  \,  \bullet  \big \}_{PB}    \,  \big \{  {\cal P}_j,  \,  \bullet  \big \}_{PB}  \,  + \, \widetilde{V} \left( \{ {\cal X}^i  \} \right)
\label{hqm}
\end{eqnarray}
The Hamiltonian flow for operators acting on ${\cal H}_{QM}$ is associated to the Hamiltonian $H_{qm}$. This flow is induced by the symplectic structure governing the canonical commutation relations of the algebra of observables ${\cal X}^i$ and $\tilde{\kappa}  \big \{  {\cal P}_i,  \,  \bullet  \big \}_{PB}$ (which  itself is a sub-algebra of the algebra in Eq.~(\ref{calops})). The resulting operator evolution respects the noncommutative Poisson structure and hence takes the form:    
\begin{eqnarray}
  \frac{d F}{d t}  &=&  \frac{\partial F}{\partial t}  +  \frac{1}{\tilde{\kappa}} [ F,  H_{qm} ]  
\end{eqnarray}
with an additional term $\frac{\partial F}{\partial t}$ for operators that carry an explicit time-dependence.  Notice that when $\tilde{\kappa} = i \, \hbar$, this is exactly   the  Heisenberg equation of motion for $F$. Using this to compute time derivatives of the quantum operators  ${\cal X}^i$ and $\tilde{\kappa}  \big \{  {\cal P}_i,  \,  \bullet  \big \}_{PB}$ gives:
\begin{eqnarray}
\tilde{\kappa}  \,  \frac{d {\cal X}^i }{d t}  \,  =  \, [ {\cal X}^i,  H_{qm} ]  \,  =  \,  \frac{ \tilde{\kappa}^2 }{ m}  \,  \big \{  {\cal P}_i,  \,  \bullet  \big \}_{PB}   \\   
\tilde{\kappa}  \,  \frac{d \, \tilde{\kappa}   \big \{  {\cal P}_i,  \,  \bullet  \big \}_{PB}  }{d t}  \,  =  \, [ \tilde{\kappa}  \big \{  {\cal P}_i,  \,  \bullet  \big \}_{PB} ,  H_{qm} ]  \,  =  \, -  \tilde{\kappa}  \,  \frac{\partial \, \widetilde{V} ( \{ {\cal X}^i \} )  }{\partial \, {\cal X}^i }   
\end{eqnarray}
Notice that these are yet again realizations of Hamilton's equations for the quantum operator algebra, and their expectation values will yield the familiar version of Ehrenfest's theorem in quantum mechanics. To see this, we apply   expectation value of the second kind to the above equations with respect to the quantum mechanical wave function $\psi_{QM}$ to get:  
\begin{eqnarray}
 \angled{ \frac{d \,  {\cal X}^i }{ d t }  }  \, = \,  \frac{1}{m}  \angled{ \tilde{\kappa}  \big \{  {\cal P}_i,  \,  \bullet  \big \}_{PB} }    \qquad \qquad    \angled{  \frac{d \,  \tilde{\kappa}  \big \{  {\cal P}_i,  \,  \bullet  \big \}_{PB} }{ d t }  }   \, = \,  -  \angled{ \frac{\partial \,  V \left( \{  {\cal X}^i \} \right) }{\partial \, {\cal X}^i }  }
\end{eqnarray}
where working in the Schr\"{o}dinger picture, the time derivative can be pulled out of the expectation value brackets $\angled{\cdots}$ to retrieve the standard form of the Ehrenfest's theorem.  
%As a corollary of this result one obtains  
%\begin{eqnarray}
%\kappa  Lie bracket \to Poisson bracket
%\end{eqnarray}
%holds in standard quantum mechanics (the proof follows [CITEEEE......]).

What we have shown above is that quantization realized as "twisting" of sections of the bundle ${\cal E}$ gives quantum mechanics from Koopman-von Neumann's classical mechanics.  At the level of the algebra Eq.~(\ref{calops}), the map Eq.~(\ref{alghalf}) identifying one sub-algebra with another is equivalent to performing a quotient on the algebra of classical operators to obtain a quantum algebra which is half of the classical one. From this perspective, the well known identification of the quantum momentum operator as a derivative $\hat{p} = - i \, \hbar \, \partial/\partial x$  is simply an instance of this quotient map. 

It is worth noting that other quantization schemes slightly different from the one we have introduced here are also plausible when working with modifications or generalizations of the Koopman-von Neumann equation. In particular, a specific  generalization of the Koopman-von Neumann equation is considered in  \cite{klein2018koopman}.   This work generalizes the Liouville operator (also suggesting an infinite family of possible generalizations based on integrability). Upon considering one such choice of a generalized Koopman-von Neumann equation, the authors apply the following quantization map to recover the time-dependent Schr\"{o}dinger equation:  
\begin{eqnarray}
{\cal P}_i \to \tilde{\kappa}  \, \big \{  {\cal P}_i,  \,  \bullet  \big \}_{PB}   \qquad  \qquad   \tilde{\kappa} \,  \big \{  {\cal X}^i,  \,  \bullet  \big \}_{PB}  \to  0
\end{eqnarray}
This form of quantization was identified as "polarization"  \cite{woodhouse1997geometric,klein2018koopman}  and was interpreted as effecting a projection of the classical wave function to configuration space, thus yielding the quantum wave function. In contrast, we have considered the standard form of the Koopman-von Neumann equation and observed that this is enough to recover the Schr\"{o}dinger equation.  Apart from the aforementioned works, yet another operator approach was pursued  in    \cite{bondar2012operational}   which was shown to yield both, the Koopman-von Neumann equation as well as the Schr\"{o}dinger equation. Even though   \cite{bondar2012operational}   do not directly delve into the issue of quantization, the algebra of operators upon which their Liouville operator and their Hamiltonian depends on shows the same feature we find here via quantization: that the quantum algebra is half of the classical algebra.

\section{Conclusions and Outlook }

\subsection*{ Summary }

In this work, we have argued that an abstract noncommutative operator algebra   defined by a noncommutative Poisson structure, a symplectic structure and a noncommutative differential structure serves as a plausible candidate for a pre-quantum theory. To the extent that no a priori assumption of a Hilbert space of states or wave functions is even necessary to start with. All those aforementioned  structures are constructions built upon our pre-quantum operator algebra, which may therefore be viewed as a foundational starting point of a quantum  theory.  This formalism was referred to as Operator Mechanics (OM). OM serves as a pre-quantum algebra from which algebraic structures relevant to real-world  classical and quantum mechanics follow. In this work, we have shown systematic constructions of  homomorphisms from  OM to classical structures, to quantum structures. In particular, the classical Koopman-von Neumann wave function as well as the quantum wave function are both consequences of this pre-quantum formalism.  

With OM, we derived the operator versions of Hamilton's equations and Newton's law of motion. These follow from the underlying noncommutative differential  and symplectic structures.  We then identified a pre-Hilbert space ${\cal H}_{pre}$ within this operator algebra, which enabled us to define a density operator based purely on  operators. This allowed for a  computation of  expectation values of operators without invoking any notion of wave functions or states.  These were referred to as expectation values of the first kind. Using this operation, expectation values of $X^i$ and $P_i$  in the large $N$ limit (of operator dimension) led to coordinates ${\cal X}^i$ and ${\cal P}_i$ which provide the spatial support for an algebra of  functions  $f \left( {\cal X},  {\cal P} \right)$ over these variables.  The coordinates $({\cal X}^i, \, {\cal P}_i)$ define a phase space and the above-mentioned functions can be thought of as sections over the phase space bundle ${\cal E}$.  The classical Koopman-von Neumann wave function  $\psi_{KvN}$ is precisely such a section of ${\cal E}$, and the space of these sections along with the norm induced from $L^2$ functions, yields the Hilbert space ${\cal H}_{KvN}$. In this sense, the origins of $\psi_{KvN}$ can be traced to the OM pre-Hilbert space ${\cal H}_{pre}$ of operators.  Successive homomorphisms then take us to the familiar space of wave functions ${\cal H}_{KvN}$ and the algebra of observables acting on ${\cal H}_{KvN}$. Along the way we also encountered the Hamilton's equations in various guises; from the operator version, to the classical commutative version, to those appearing in Ehrenfest's theorems; suggesting a certain functorial property of these equations that is preserved across algebraic structures. 

An analogous picture emerges for the quantum mechanical wave function  $\psi_{QM}$ and its familiar home in ${\cal H}_{QM}$, with the caveat that the former depends only on ${\cal X}$ and not on ${\cal P}$.  We showed that the  difference between  $\psi_{KvN}$ and $\psi_{QM}$ comes from a quantization scheme interpreted as  "twisting" of sections over  ${\cal E}$. Alternatively, this twisting can also be viewed as the quantum algebra of observables being half the classical one.  Using such a quantization map, we showed that the Schr\"{o}dinger equation itself can be obtained starting from the Koopman-von Neumann equation. What all this suggests is that neither the Schr\"{o}dinger equation nor the quantum mechanical wave function  are fundamental structures. All these are simply built upon a pre-quantum operator algebra and ensuing algebraic homomorphisms.  

Below we summarize the key algebraic structures involved in OM, classical mechanics and quantum mechanics. This comparison also serves to illustrate the role of OM as a pre-quantum operator algebra from which real-world  classical and quantum mechanics follow upon loosening specific facets of its algebraic structure: 

\begin{itemize}

\item Operator Mechanics (OM) is formalized using a noncommutative Poisson structure, a symplectic structure, and a noncommutative differential structure.

\item Classical Mechanics on the other hand is formalized using a commutative Poisson structure, a symplectic structure, and a commutative differential structure from standard analysis.  The transition from OM to classical mechanics involves homomorphisms from the noncommutative Poisson structure of OM to a commutative Poisson structure, and from noncommutative calculus to the classical one.  

\item Quantum Mechanics is also formalized using a noncommutative Poisson structure and a symplectic structure. However, the differential structure realized  here is, in part, that of classical calculus inherited from a projection of classical phase space onto configuration space. While the transition from OM to quantum mechanics retains noncommutativity of the Poisson structure, it partly loses OM's noncommutative differential structure. The latter is because the differential operator $\partial / \partial {\cal X}^i$ in quantum mechanics is the ordinary derivative of configuration space (in contrast to $\partial / \partial X^i$, the noncommutative derivative in OM).  The Schr\"{o}dinger operator in quantum mechanics is expressed in terms of this differential operator from classical  analysis. 

\end{itemize}

Note that the above-mentioned noncommutative differential structure of OM strongly suggests an  underlying noncommutative pre-geometry  associated to this operator algebra. This noncommutative pre-geometry would presumably be required for investigating  structures relevant to classical and quantum theories of gravity.

%Here, the wave function $\angled{\phi(X)}$ carries a dependence only on the eigenvalues of  the $X^i$.  A key step that enabled the derivation above was that we were able to  bootstrap all the non-linear dependence of $\angled{X}$ in $\angled{\phi(X)}$ into operators $\phi(X)$ linear in $X$. The problem then   reduces to parametrizing $\phi(X)$ in terms of undetermined functions $c_i (\angled{X})$ and solving the operator equation of motion Eq.~(\ref{dt}). To solve this operator equation, we invoked operator expectations values on both sides of the equation, and once again the above bootstrapping came to the rescue.  The result of this procedure being that the operator evolution equation induced by the Hamiltonian flow led directly to the inhomogeneous Schr\"{o}dinger equation for $\angled{\phi(X)}$ in the large $N$ limit (where $N$ corresponds to the dimension of the operators). In our set-up, Eq.~(\ref{dt}) was associated to the Hamiltonian flow induced by the symplectic structure. Notice that it was precisely Eq.~(\ref{dt}) that also led to the operator form of Hamilton's equations (along with the canonical commutation relations). And now, a different guise of this very same equation leads to the Schr\"{o}dinger equation (with the choice of $\kappa$ being $- i \, \hbar$). More generally, one may also consider obtaining   Eq.~(\ref{dt})  from the saddle-point of an operator action. In any case, solving for a generic wave function would still likely require some form of bootstrapping  as discussed above. 

\subsection*{ A Route to Quantum Field Theories Using Matrix Models}
 
Our work here suggests several possible routes of extension. The obvious ones being formalizing OM for relativistic quantum mechanics and deriving the Klein-Gordon equation. Another interesting possibility would be to compute sub-leading $1/N$ corrections to the Schr\"{o}dinger equation. Yet another exciting extension would be to formalize an operator algebraic field theory using matrix models. Let us outline this prospective application of OM.

The objective in this case would be to construct an operator algebraic formalization of quantum field theories without assuming an a priori classical geometry or Hilbert space. We present the outline of the idea and will soon report details in an upcoming work. A plausible extension of OM to a field theory may be realized by considering instead, an ensemble of Hermitian operators $X^i$ and $P_i$ defined with respect to an appropriate measure. In fact, a precise realization of such an ensemble and its associated measure can be obtained by representing $X^i$ and $P_i$ as $N \times N$ Hermitian matrices.  We propose that the path integral of this field theory is a special type of matrix model. In fact, this is what we will call a `Constrained One-dimensional  $2n$-Matrix Model'. The constraints refer to the OM canonical conjugation relations between the $2n$ matrix operators $X^i$ and $P_i$. The action of this one-dimensional matrix model for the non-relativistic case can be expressed as:
\begin{eqnarray}
S_N [\phi(X)] = N \, Tr  \int dt \left( \frac{m}{2}  \dot{\phi}(X)^2  \, - \, V \left( \phi(X) \right)  \right)
\end{eqnarray}
where the OM constraints corresponding to the canonical conjugation relations and subsequent partial derivatives have been used to cast the above action $S_N$ in terms of $\phi(X)$ and $\dot{\phi}(X)$. The integral above is a formal antiderivative defined using the operator inverse of Eq.~(\ref{dt}). As indicated,  the constraints due to the canonical commutation relations  help remove the $\chi(P)$ dependence from the action, thus simplifying the problem.      
The operators $\phi(X)$ are chosen as elements of the pre-Hilbert space ${\cal H}_{pre}$ satisfying Eq.~(\ref{nm}). The partition sum for this action is the following matrix integral:  
\begin{eqnarray}
{\cal Z}_N  =   \int {\cal D} \phi(X) \, {\cal D} \chi(P)  \, \, e^{\frac{i}{\hbar}  S_N [\phi(X)]}  
\end{eqnarray}
The measure ${\cal D} \phi(X) \, {\cal D} \chi(P)$  runs over Hermitian matrices $X^i$ and $P_i$, as well as the space of coefficients of the relevant ${\cal R}$-module.  

A special case of the above matrix model, where one sets $\phi(X) \to X^i$, would give precisely the standard one-dimensional matrix model in $X^i$ (see \cite{anninos2020notes}). The above path integral is a natural extension of OM to a field theoretic setting. Presumably the large $N$ limit of this matrix model  approximates certain classes of low-energy scattering amplitudes of a scalar field theory, and it's $1/N$ corrections might help open a window into non-perturbative contributions in a manner similar to previously known correspondences of matrix models to topological field theories \cite{dijkgraaf2002matrix}.  Let us mention another closely related matrix dynamics approach which also seeks to formulate quantum field theories and ultimately gravity  without classical spacetimes is Adler's `trace dynamics'  \cite{adler2004quantum,singh2021trace}. This is based on Grassmann-valued matrix operators rather than Hermitian matrices.  Like OM, this program also posits noncommutativity at the fundamental scale. It will be interesting to explore connections of OM to trace dynamics in follow-up work.

\subsection*{ Connections to Noncommutative Geometry}

As suggested earlier, a potential avenue for future exploration would be possible connections to  Noncommutative Geometry (NCG) \cite{connes1985non}.  Even though the objectives of our work in this paper did not concern geometric considerations,   the OM formalism shares  conceptual similarities to Connes' noncommutative geometry, in that differential operators are defined as commutators with respect to a compact operator. The central thesis of Connes' work seeks to generalizes the Gelfand-Naimark Theorem and posits that geometric data can be fully constructible from the ${\cal C}^\star$ algebra of noncommutative operators over a Hilbert space. Unlike NCG, in OM we assume less than the full spectral triple (starting with only the operator algebra and a designated Hamiltonian), and obtain the Hilbert space of states as a by-product of the algebra. In a sense, this means OM is agnostic to the choice of representations of the algebra, which can have interesting implications for exploring the moduli space of possible geometries.   

Besides the NCG program, other realizations of noncommutative geometry include `Sub-Riemannian Geometries' associated to Carnot-Caratheodory groups  \cite{buliga2002symplectic,buliga2007dilatation}. The Heisenberg group being a special case of a Carnot-Caratheodory group, potentially suggests a sub-Riemannian geometry for modular spaces associated to representations of the Heisenberg algebra discussed in \cite{freidel2016quantum}.

\subsection*{ Relation to Quantum Foundations and Spaces from Entanglement }

In the current work, we have not said anything about the various interpretations of quantum mechanics or the measurement problem. Rather the focus of our work here has been on elucidating formal structures that provide the foundations for spaces upon which contemporary formulations of classical and quantum mechanics are described. It is hoped that such structural foundations may shed insights on on-going debates concerning ontological questions such as wave function realism, fundamentality of Hilbert spaces and emergence of classical spaces via entanglement. For now, let us comment only on these three issues. 

Recall that wave functions are not exclusive to quantum mechanics. This was already demonstrated since the work of Koopman and von Neumann   in 1932   \cite{koopman1931hamiltonian,neumann1932operatorenmethode}.   What we have emphasized in our work here is that both, the classical and quantum wave functions $\psi_{KvN/QM}$ are sections of the same phase space bundle ${\cal E}$. The difference between the two being a geometric twist of sections, otherwise interpreted as a quantization scheme on the algebra of observables. Furthermore, the Hilbert space ${\cal H}_{KvN/QM}$ in which these wave functions reside is again not exclusive to quantum mechanics.  In this view, neither the wave functions $\psi_{KvN/QM}$ nor their associated  Hilbert spaces ${\cal H}_{KvN/QM}$ are fundamental structures. Rather, these are merely "front-ends" supported on a tower of structures built upon a pre-quantum algebra. Both $\psi_{KvN}$ and $\psi_{QM}$  belong to the algebra of  functions over phase space (${\cal X}$ and ${\cal P}$) respectively configuration space (${\cal X}$) variables, which themselves are obtained as  expectation values of the first kind associated to operators $X$ and $P$ of the OM pre-quantum algebra. The familiar Hilbert spaces ${\cal H}_{KvN}$ and ${\cal H}_{QM}$ arise from the space of these sections of the phase space bundle ${\cal E}$ along with the norm induced from $L^2$ functions. Given that the operator expectation values of the first kind originate from a pre-Hilbert space ${\cal H}_{pre}$ of operators in OM, the origins of both, the classical and quantum wave functions $\psi_{KvN}$ and $\psi_{QM}$ can be traced to this pre-Hilbert structure and successive algebraic homomorphisms. Under this unified perspective, the Koopman-von Neumann's and Schr\"{o}dinger's equation, both appear on an equal footing. We showed how the latter may be viewed as a quantization of the former. 

What insights does our framework provide on emergence of conventional  spaces upon which physics is "operationalized"? In particular, we noted that operator expectation values of the first kind defined in OM, led to a  characterization of phase space involving variables ${\cal X}$ and ${\cal P}$ as coordinates. But this is just a set of "points" (${\cal X}$,  ${\cal P}$).  How do topological or spatial structures such as a metric get endowed upon this set? One possible route is via entanglements of blocks or modes of the operators $X^i$ and $P_i$. To see how this works, let us recall the density operator in ${\mathcal H}_{pre}$ expressed in the basis of the algebra ${\cal L}$:
\begin{eqnarray}
\hat{\rho}  =  \frac{1}{\kappa^2 } \sum_i^n  w_i  \, | \, \overline{X^i} \, ]  \,   [ \, X^i \, |
\end{eqnarray}
For operators of dimension $N$ (as before, thinking of them as $N \times N$ matrices), $\hat{\rho}$ is an $N \times N$ matrix. To quantify the entanglement entropy between operator blocks, we need a reduced density matrix $\hat{\rho}_{M}$ corresponding to an $M < N$ dimensional eigenmode of the matrix operators. This can be expressed by way of a partial trace over the complementary $N - M$-dimensional operator modes $\overline{M}$:
\begin{eqnarray}
\hat{\rho}_{M} = Tr_{\overline{M}} \, \hat{\rho}
\end{eqnarray}
where the partial trace here involves $\overline{M}$-dimensional operator eigenmodes in ${\mathcal H}_{pre}$  to obtain an $M \times M$ reduced density matrix. Using this operator, the entanglement entropy between operator eigenmodes $M$ and $\overline{M}$ is given by:
\begin{eqnarray}
S_M = -  \, Tr_{M} \, \left( \hat{\rho}_{M} \, log \,  \hat{\rho}_{M}  \right) 
\end{eqnarray}
Given $S_M$, one readily obtains the mutual information $I (M_1 : M_2)$
 between any two blocks $M_1$ and $M_2$. While it is known that mutual information itself is not a metric, using mutual information and entropy one can define several distance functions between blocks $M_1$ and $M_2$ such as: 
\begin{eqnarray}
d (M_1, M_2) = S_{M_1, M_2} - I (M_1 : M_2)
\end{eqnarray} 
known as Variation of Information \cite{meilua2003comparing}; or 
\begin{eqnarray}
d (M_1, M_2) = 1 -  \frac{ I (M_1 : M_2) }{ S_{M_1, M_2} }
\end{eqnarray} 
known as the Rajski distance \cite{rajski1961metric}, which is closely related to the Jaccard distance from a set-theoretic interpretation.  In our particular case, by considering one-dimensional operator blocks $M$, expectation values of the first kind corresponding to operators $X^i$ and $P_i$ led to $n \times N$ points  
$\left( \angled{X^i}, \angled{P_i} \right)$ denoting a phase space. Now the above definitions of distance endow this set with the structure of a metric space. And in the large $N$ limit of operator dimension, we obtain the continuum limit of this space with coordinates (${\cal X}$,  ${\cal P}$). This is how we obtain phase space from entanglement starting from eigenmodes of operators in the pre-Hilbert space ${\cal H}_{pre}$ of  OM. 

The entanglement entropies of the operator blocks defined above, do not depend on states of any system, but follow exclusively from the operator algebra. A similar argument proposing emergence of spaces from entanglement was made in \cite{carroll2019mad,carroll2022reality}, with the important difference that here we do not even need states, wave functions or the Hilbert space of standard quantum mechanics. In light of the program being pursued in  \cite{carroll2022reality},  as to `What might be the minimal starting point of quantum mechanics?', our investigations here show that the rabbit-hole of minimalism seems  to run much deeper that was previously anticipated.

\subsection*{ Connections to Geometry from Graphical and Category-Theoretic Models}

Operator algebraic approaches such as OM and its extensions may also be useful tools for formally bridging discrete models of quantum geometry to their continuum limits, and in the process systematically extract non-perturbative corrections to continuum observables stemming from the underlying discretization. This is because by themselves operator algebras are neither continuous nor discrete. The topology that the spectrum of operators inherits, depends on the representation one works with, and this can be either continuous or discrete with formal techniques to understand transitions from one regime to another. 

In particular, operator algebras similar to the ones we have discussed here may have potential foundational implications for investigating  geometry from random matrix models  \cite{arsiwalla2006phase,arsiwallasupersymmetric,eynard2015random},  emergent spacetime from entanglement  \cite{van2010building,swingle2018spacetime},  geometric structures and limiting behaviors of tensor networks \cite{evenbly2011tensor},   category-theoretic approaches to quantum mechanics    \cite{coecke2018picturing},  and combinatorial models of spacetime based on rewriting systems such as the Wolfram model   
\cite{Wolfram2002a,Wolfram2020,arsiwalla2021homotopy,arsiwalla2021pregeometric}.  Since we have already mentioned about connections to matrix models and emergent spaces from entanglement earlier, let us now briefly comment on how operator algebras may serve as useful tools for probing (emergent) geometry from graphical and combinatorial models. A key feature common to the latter are that they are based on diagrammatic calculi (as examples of diagrammatic reasoning  see \cite{kauffmans2005knot,Coecke2009a,evenbly2011tensor,coecke2018picturing,Wolfram2020})   that offer flexible theorem proving and formal interpretability techniques including diagrammatic reasoning and categorical semantics.  Applications of  diagrammatic approaches include  tensor networks and a host of other diagrammatically representable operator calculi used in the investigation of holographic dualities  \cite{swingle2018spacetime,arsiwalla2011degenerate}, string diagrams used in categorical quantum mechanics and quantum circuit simplification \cite{Coecke2009a,coecke2018picturing}, and hypergraph rewriting systems used in discrete models of spacetime and pregeometric spaces appearing in the Wolfram model \cite{arsiwalla2021pregeometric}.  Many of these approaches find a common home in category (and higher category) theory.  

For instance, consider formal rewriting systems as the ones espoused by the Wolfram model.  A rewriting rule such as 
\begin{eqnarray}
{\cal G}_1 \to {\cal G}_2
\end{eqnarray} 
involving graphs or hypergraphs, when applied to an initial state generates a multi-history evolution network, respecting local causal structure. This is the Wolfram model `Multiway System'. Rewriting rules are unary operations applied on graphs, character strings or other operators. This resulting evolution sequences are formally comparable to those generated by the operator derivative Eq.~(\ref{dt}). The latter also suggests a potential graphical operator playing the role of a Hamiltonian, which would generalize unary rewriting to $n$-ary operations involving a sequence of graphs (or hypergraphs)  
\begin{eqnarray}
{\cal G}_1 \times \cdots \times  {\cal G}_n \to {\cal G}_{n+1} \times \cdots \times  {\cal G}_m
\end{eqnarray} 
as higher-order rewriting rules captured via an operator algebra of graph actions. Furthermore, multiway systems are natural constructions to express tensor networks of categorical quantum mechanics and ZX-calculus as was shown  in \cite{Gorard2020c,Gorard2021a,Gorard2021b}.  In \cite{arsiwalla2021homotopy,arsiwalla2021pregeometric} this categorical formulation of multiway systems was extended $n$-fold categories and homotopy types for the purpose of constructing synthetic pregeometric spaces. With higher-order rewriting rules, what was shown in that work was that such a limiting multiway system was formally identifiable with the $\infty$-groupoid. This nexus of higher category theory, tensor networks and homotopy type theory suggest new ways to probe synthetic geometry and pregeometric structures as higher categorical constructions.

%comment on taxonomy of possible theories from spectral doubles, etc

%comment on geometric QM here????

%extra STUFF - MM as bridge between spin models and dual QFTs
% can square of length n be partitioned by rectangles of length less than n and >1? what about analogue for hypercubes? - relation to impossible pythagorean triples and possible link to fermat's last

%notes on multiways as algebra generators and link to OA

%\clearpage
%\section*{Appendix}
%\appendix
%
%\section{Additional Notes}
%\label{sec:section5}

\section*{Acknowledgments}

We would like to thank Stephen Wolfram for his encouragement and useful suggestions. We gratefully acknowledge Marius Buliga and Djordje Minic for useful comments and feedback on the manuscript.

\bibliographystyle{eptcs}
\bibliography{hottrefs.bib}

\end{document}